# PSC: A Pattern-Based Temporal and Spatial Crowdsourcing Platform to Improve Performance, Reliability, and Privacy

Faranak Davoodi
University of Southern California
fdavoodi@usc.edu

Intelligent buoy Networks, Inc.
faranak.davoodi@intelligentbuoys.com

## ABSTRACT

In this paper, we study a novel spatial crowdsourcing system where the workers' time availabilities and their spatial locations are known a priori.  Consequently, the tasks assignment to workers is performed not only based on the current location of the human workers and the tasks available in the region, but also based on the availability of the workers during the specific times that a given task should be accepted, processed, and completed. Having the system determine the daily pattern of the workers (either by pre-defined questionnaires when the workers register, or by archiving data from the worker's mobile devices, or by on-the-road and real-time entered status data) eliminates many unsuccessful task assignments and therefore significantly increases the efficiency of the system. In the original Spatial Crowdsourcing (SC) framework, the SC-server optimizes the task assignment locally at every instance of time and whenever a new task, or a new worker, enters the system. Our new framework (PSC), on the other hand, allows the users to enter their daily routine, and temporal, spatial, and availability patterns a priori. This makes the system much more stable and pattern-opportunistic. The PSC servers can focus on receiving and archiving new entries (e.g., workers, tasks, and their criteria) during busy times (e.g., when there are many new entries in the system), and can focus on optimization and computations during quiet times (e.g., when there are fewer new entries in the system). Having the task optimization process happen during quiet times, and when there are few changes to the system, makes the performance more stable and reliable.  It also allows the PSC system to have a global view of the system and perform global optimizations to improve the performance. The details of the PSC architecture will be described, including its novel performance and query methods. Comparisons to previous SC architectures will be made and the advantages of PSC highlighted.

## Categories and Subject Descriptors

H.2.8 [Database Management]: Database Applications— Spatial databases and GIS

## 1. INTRODUCTION

A Spatial Crowdsourcing (SC) architecture for human workers was introduced in [1] to be a platform for the blooming crowdsourcing services [2-3] which exploit the recently unprecedented abundance of available multiplatform (e.g., smartphones, laptops, desktops, TVs) interconnected social networks and on-line businesses. The workers involve large crowds of ordinary people capable of requesting or providing various services (e.g., taking pictures, performing sensory acquisitions, purchasing and delivery) in certain locations and in exchange for certain benefits (e.g., monetary, humanitarian, environmental). A major challenge of the SC-Architecture, as mentioned in [1], is that "the SC-server only has a local view of the available tasks and workers at any instance of time. This means that a global optimal assignment is not feasible. Instead, the SC-server tries to optimize the task assignment locally at every instance of time. The SC-server does this by utilizing the spatial information that workers share during their task inquiries [1]".

Conversely, the SC-server performs queries based only on the spatial information of the workers and the tasks, and without considering the *availability* of the worker in a location close to the task and during the time period required for the task to be started, processed and completed. In real world applications, not considering these may potentially cause problems for the system, as it may not assign the most suitable workers to the tasks, viz., workers who would most likely accept, perform and complete the task in time and before the deadline.

We will demonstrate the importance of the above consideration using two simple examples. The first is a task that needs to be finished in 30 minutes (e.g., the time that it might take to deliver a bouquet of flowers from the origin location to a given destination in normal traffic). Assume that there are 2 workers A and B, located in the proximity of the task. Worker A is 3 miles away from the task and worker B is only 1 mile away. Worker B has some prior obligations (e.g., needs to leave his current location, immediately, in order to get to his work in time,). However, worker A will be available in the area for the entire day (for example, the worker could be a retired person or a stay-home mom living in the area). A SC-query which only considers the Euclidean distance between the location of the workers and the given task will pick worker B. This is obviously an incorrect selection. In contrary, the PSC-query will not have this problem, since it considers both the spatial distance between the workers and the tasks, as well as the period of time that a worker is available in a certain location, and the amount of time required to complete the task.

Example 2 involves an area considered to be of high entropy (i.e., many workers are located there very often) as defined in [1]. A government building, or a company in a city's downtown, are such locations. Therefore, the SC-system assumes that any task given in the proximity of a high entropy location should be accepted and completed quickly and successfully. However, without considering the availability of the workers and their daily patterns, this logic might be misleading. If the location is a government building, for example, it is likely that most of the workers are government employees and therefore by and large not willing to leave their job in the middle of a working day to perform an "extra-pay task" for a bit of money (such as taking a picture of a location 2 miles away from work). Clearly, most people would not go through the hassle of taking their car, leaving work, driving 2 miles to take a picture, driving back, finding a parking spot, and getting back to work while worrying about the consequences of being caught by their supervisors.

In the PSC architecture we have defined a platform that we believe can overcome the challenges mentioned above and will be a more comprehensive crowdsourcing system to match real-life situations. In the second example above, the PSC-workers who are also the employees in the government building could simply mark themselves as "unavailable" to do any PSC-task for all hours that they are at their daily-work. This way, the PSC-sever can simply eliminate several unavailable employees from its queries. This will save a lot of overhead and CPU-time, since wrong selections, and the following rejections of the selected workers, will result in the system repeatedly performing the query, assignment and rejection (over and over) until the right workers are spotted.

We should mention that we may also consider a situation where a "busy" worker might accept a task, as an exception, if there is a high reward that is given for performing the task. A questionnaire at the time the worker is registering could verify the details of such exceptional situations when the worker agrees to perform a task when he/she usually marks his/her status as "busy and unavailable"

***This paragraph needs to be rewritten after the rest of the paper is finalized.** In sections 2.1 to 2.10 we will define a PSC-platform, consisting of comprehensive and novel attributes for workers, tasks, task-owners, patterns, regions, costs, trustworthiness-scores, PSC-performance, etc. For each additional and enhanced PSC-attribute, we will bring examples and numerical analysis to show how they are different from prior work on the SC-platform and how they can improve the performance of the system and better match real-world applications. In section YY.1, we will introduce an algorithm, which shows how to use the PSC-attributes defined from XX.1 to XX.10, to perform a PSC-query. In section YY.2, we will perform numerical simulations to highlight the improvement of the PSC-system compared to the original SC-platform introduced in [1], [4], [5-7].*

## 2. PSC platform

In this section we go through the detail attributes of the PSC system. We believe that the novel PSC platform described here will be capable of providing a real-life matching platform for on-line crowdsourcing businesses where there could be up to millions of people, visitors and users. The PSC platform is designed to give flexibility to the crowdsourcing servers to focus on receiving and archiving numerous new and updated users, and their profiles and criteria, during the servers' busy-poking-times (e.g., during the day), and to perform thousands of optimized queries and task assignments during the servers' quiet times. Moreover, the system's novel definitions of patterns, regions, costs, etc., will create a win-win-win situation to get the most benefit for the workers, the task-owners, as well as the PSC service providers.

## 2.1 PSC-Performance:

We begin with the definition of the performance of the system.

***Definition 1: (System Performance)*** *The performance of the PSC system is defined as the total number of task requests that are performed and completed by the workers, as defined by the task-owners, during a given amount of time.*

This definition of performance is from the vantage point of an external observer to the system who is measuring the number of completed tasks per unit time. Clearly, for a task to be completed, it should be 1) assigned to a worker, 2) accepted by a worker, and 3) performed and completed to the satisfaction of the task owner.

An alternative measure of performance is the fraction of completed tasks, relative to the total number of tasks submitted to the system, during a given amount of time. This measure of performance prefers lighter system loads (i.e. if the number of submitted tasks is small, there is a better chance that a large fraction of them will be satisfactorily completed). For this reason, we prefer the performance measure in Definition 1 since it measures the "total throughput" of the system.

We should mention that in its performance evaluations [1] considers two different performance metrics. The first is the total number of assigned tasks. We prefer our metric to this since not all assigned tasks are accepted and completed, and that, at the end of the day, what one cares about is the number of *completed* tasks. The second is the average travel cost for a worker to perform a spatial task, where the travel cost is measured by the distance the worker needs to traverse. Unlike the metric in Definition 1, this is an "internal" metric not measurable by an observer who can see the number of tasks completed. Furthermore, any assignment scheme that maximizes the performance according to Definition 1 will consequently be (roughly) minimizing the average distance traversed by the workers, since if a task is assigned to a worker at a great distance it is highly likely that the worker will turn it down and not complete it.

## 2.2 PSC-Task-Owner

The task owner can submit tasks to the system. More formally:

***Definition 2: (PSC-Task-Owner or PTO)*** *The PSC-Task-Owner (PTO) is a user of the PSC platform that is registered in the system and can submit some tasks to the PSC-system for execution.*

When a PTO registers the PSC-system provides a questionnaire to the PTO in order to create a profile. Besides the personal information (e.g., full name, address, payment information), the PTOs are encouraged to (ahead of time) provide information about the patterns of the tasks that they plan to submit for execution. For example, the type of the tasks, the approximate location and time, the typical payment that the PTO offers (either based on the time spent to perform a task, or the on the successful execution), etc. Moreover, the system will give higher priority of service to those PTOs who provide the system with details of their requested tasks in advance. Moreover, the system will make sure that the PTOs with better reputation of payment (paying in time, and paying appropriately or even higher) would get better service (in order to encourage them to stay with the PSC-system-provider). To improve the performance, the PSC-system can assign a PTO-Priority-Score to each PTO as described below in section 2.2.1.

### 2.2.1 PTO-Priority

***Definition 3: (PTO Priority)*** *The system will associate to each PTO a priority score, $0 \leq PTOPriority(PTO) \leq 1$, which reflects the priority the system gives to tasks submitted by the PTO when it assigns tasks. $PTOPriority(PTO)$ depends on the PTO's behavior in the system and can change accordingly.*

The introduction of $PTOPriority(PTO)$ is to help the system make more efficient decisions. As the name suggests, the system will give higher priority to PTO's with higher priority scores when assigning tasks. A few suggestions for when a PTO's priority should be set higher (and their effects on the performance of the system) are as follows.

A PTO who does not meet his financial obligations on time for the tasks that have been completed successfully by the workers receives a lower priority-score. If the workers are not paid on time after performing a task, they will more likely reject new tasks assigned to them. Any rejection of a task by the workers will negatively affect the system performance as given in Definition 1; therefore, by giving less chances, or perhaps even eliminating PTOs with bad-priority-scores, one could increase the performance of the system. Conversely, PTOs who pay on time will get higher priority scores. Since, workers will be more encouraged to work for a PTO whose reputation is higher and who pays on time, the number of worker rejections and incomplete tasks will be reduced.

Also, those PTOs who bring their tasks ahead of time, and bring several of them, will be given higher priority. The more number of tasks that are given to the server ahead of time, the higher the system's performance will be. The reason is that the PSC-server can have the necessary information (location, award, duration, etc.) to perform the query and optimal task assignment ahead of time and therefore will have enough time and information to perform global optimization. However, those customers who bring their tasks to the system in short notice (the tasks need to be started and completed in a short matter of time from its request by the PTO) will get a lower priority score. Since handling their tasks in real-time will only allow the server to perform a very limited local search, the quality of the assignment returned could not be as good as when the server spends enough time for the optimization. PTOs who submit their tasks in short notice can compensate for their low priority score by increasing the award for the task they request. We shall comment more on these when we describe the assignment algorithm.

## 2.3 PSC-Task

***Definition 4: (PSC-Task or PT)*** *A PSC-Task (henceforth PT) is what a PTO requests for execution to the PSC-system. Along with the task description and category, the PTO will submit several sub-attributes, which helps to optimize the PSC assignment.*

Below we will go over some of the above attributes and will show how they can affect the system performance. In section 3, we will demonstrate in detail how the PT sub-attributes mentioned here can be used in the PSC assignment algorithm.

### 2.3.1 PT-id
$id(PT)$ is a unique natural number assigned to each new task entered into the system.

### 2.3.2 PT-Description

For any task PT, the $Description(PT)$ returns a text describing the task in words so that it is comprehensible to the worker whom the task is assigned to. For example, a task description that reads, "pick up a flower bouquet from shop A and deliver it to address B". The text for the task description must be entered by the task owner at the time he/she submits the task to the PSC system.

The PSC-platform might provide a text box along with various drop-down lists, and interaction maps (e.g. Google map) to help the PTO to specify important information about the requested task (such as region, task category, reward, etc.) which is required for an efficient task assignment by the system.

While the task description is primarily for the use of the worker, the PSC system might have sophisticated text-parsing and understanding software in order to retrieve important information about the PSC's requested task (e.g. task category, location, reward) from the task description itself. For example, a task description such as "take a picture of the Hollywood Sign at sunset", could be parsed to retrieve information such as "taking a picture" for the task category, "regions where the Hollywood sign is visible" for the task region, and "a window of a certain duration around sunset" for the task start time and expiration time.

### 2.3.3 PT-Category

$TaskCat(PT) \in S_{cat}$ returns the category of the task, where $S_{cat} = \{c_1, c_2, ..., c_m\}$ is a finite set of categories. Examples of categories could be "delivery", "taking a picture", "picking up someone", etc. Each PSC task category has a priority $TaskCatPriority(c) \in [0,1]$ that helps the system decide about the order of the tasks that need to be processed and assigned. A task with a higher $TaskCat(PT)$ should be processed first. For example, a task category such as driving someone to the airport has a higher priority than a task category such as delivering a flower. Therefore, the PSC-server should try to assign an appropriate worker to drive the person to the airport, as soon as possible, and before handling lower priority tasks such as delivery of the flower.

The actual definition of the priority of a task depends not only on the priority of the category the task belongs to, i.e., $TaskCat(PT)$, but also on the task owner's priority for the system $PTOPriority(PTO)$, and the priority that a task owner might manually enter for the submitted requested task, which we shall refer to as $PTEnteredPriorityScore(PT)$. More details on how to calculate the actual priority of the task will be described in section 2.3.8. below.

PSC-system will assign an estimated and tunable reward value to each task category $TaskCatReward(PT)$ such that all the tasks in the same task-category would be rewarded almost the same. We will demonstrate in the section 2.3.9 how we use the task category to calculate the PSC-worker's task reward.

Moreover, the task category could be used in order to verify how well various workers could be in performing some specific task categories. For example a person might always reject a task in the delivery category (perhaps because he or she does not drive or does not have a car). Identifying and archiving those task categories that a worker is not good at will help the PSC-system ensure that is does not assign them such workers. Consequently, this strategy will decrease the possibility of the rejection of assigned tasks (which were made inappropriately) and therefore increase the performance of the system. In the section 2.4.5, we will demonstrate how $TaskCat(PT)$ is used to compute the trustworthy score of each worker.

### 2.3.4 PT- Start Time

$StartTime(PT) = [t_1, t_2]$ is a time interval indicating the acceptable range of times for starting the PT. It is entered by the PTO when the PT request is put into the system. If the lower bound $t_1$ is not entered, it means that the task should start no later than $t_2$; if $t_2$ is not entered, it means that the task should start no earlier than $t_1$; if neither extremes are entered, it means that the start time of the task is not important.

As mentioned earlier, a PTO might submit the task a few weeks ahead of time. The PSC-system encourages such advance submissions, since it can improve the efficiency of the system. If $StartTime(PT)$ is of very short-notice, the system might not start immediately, if it is busy with other higher priority tasks.

### 2.3.5 PT-Duration Time

$DurationTime(PT) = t$ is a real number indicating the expected time the PT takes. This is only a function of the task and consequently does not include the time it may take for a worker to get to the location where the task needs to be performed. It is entered by the PTO at the time the PT is submitted to the system.

In more sophisticated systems, $DurationTime(PT)$ could be time-dependent. For example, if the task is the delivery of a bouquet of flowers from point A to point B, the duration time may change depending on the traffic patterns during different times of the day (such as rush hour). It is reasonable to assume that the system might have access to live traffic websites and information and could tune the $DurationTime(PT)$ accordingly.

There has been no indication of the duration of the task in the previous works [1], [4-7]. However, the task duration information could be extremely vital for the system to decide which workers are available to start and finish the task before they need to go after their other daily obligations. In previous systems, a task that might take 2 hours could have been assigned to a worker that needs to leave the area in 30 minutes. This would have caused the worker to reject the task and therefore hurt the performance of the system. However, in the PSC, the system will be aware of the duration of the task and will make sure to only assign task to workers who are available to perform the task for that time period.

### 2.3.6 PT-Expiration Time

$ExpirationTime(PT) = t$ is a time and date entered by the PTO such that, if the task is not completed by then, the PTO has no interest in the completion of the task and has no obligation to pay for it. It is essentially the deadline of execution of the task.

The definition of $ExpirationTime(PT)$ is very similar to the one described for the SC-platform. As explained in [1], an "expiration time [1]" could be assigned to each task and "the expiration time of every task can be utilized as a tiebreaker in the assignment process [1]". Note that the expiration time is different with the duration of a task. The expiration time is set by each task-owner when the task is submitted. We need to mention that in [1], even when the expiration time was defined as an attribute for the SC-task, it was not used in any of the formulas or algorithms, including the task-assignments. Accordingly, in the definition of the updated SC-task in [4], the definition of a task no longer included the task expiration.

A more flexible PSC-system might try to consult a worker whose task is finished with only a short delay after its set expiration time, in order to negotiate the completion of the task in exchange for a slightly discounted award (the PTO would be able to pay less than what it was supposed to pay in the first place). Conversely, the PTOs might enter their level of tolerance for performing a task late (e.g., a PTO might agree to pay half what it is supposed to pay for a timely completion of the task if the task is done within 24 hours of its original expiration time), in advance and when submitting the task. Of course, having the PSC-system being flexible as mentioned here would increase the performance of the system since more tasks will be considered to be completed and successful. We may study such flexible expiration times and their effect on PSC system performance in our future work.

### 2.3.7 PT-Location and Region

The SC-location of the task, as described in [1] and [4], is said to be only the exact address or geo-location of each task in the system that needs to be performed. However, in the PSC-system, the tasks have the flexibility of being assigned to an exact location, or to a region.

$TaskLocationRegion(PT) \subset R^2$ is a region that is specified by the task-owner. A suitable format for entering this information should be designed. The format should allow the flexibility of assigning exact locations (based on exact address or coordinates), assigning areas such as those inside a given rectangle on the map, or assigning entire neighborhoods, etc. If a PTO specifies a region for a task, it means that the specific task should be executed in the entire region. An example the task could be distributing 100 flyers of a business (e.g., a new restaurant) in the region and in a radius of 1 mile from the business. Another example could be to take a picture of a whale in a region of 100 square miles from Santa Monica Beach, CA towards the south. Having a region instead of exact locations, gives much more flexibility to the system, and therefore can improve the performance compared to the SC-platform where tasks are given exact locations. In the former example, if the system can only assign an exact location, it would have to try numerous queries and false assignments (to workers who may not be willing to go to the exact address) before finding the right workers ready to go to the exact location and distribute the flyers. However, when a wider region is assigned to a task, there are many more opportunities (every single address in the region could be a correct query for any worker who can match it). Therefore, the probability of finding workers who can accept the task in time will be higher, improving the performance of the PSC-system. The same is true in the example of taking a picture of a whale, if a wide region in the ocean, instead of an exact one, is specified. Any worker, inside or near the region could be a match to perform the task.

Another application for specifying a region for a task is when the reasonable typical reward to perform the task needs to be calculated and proposed to the workers and task-owners. For example, performing a task in some parts of a city with higher crime rates, or one under some hazardous condition (e.g., flood, earthquake, etc.), should cost much more than regions where everything is safe and sound. The same is true about regions with much fewer numbers of available workers. For example, consider a task-owner requesting to have a picture of an ice-ridge in the Arctic, or of some location in the Sonoran Desert, CA, where there is a low possibility of finding available and appropriate workers to perform the task. If a higher reward is not given to perform tasks in such "difficult" regions, then the probability of the task being rejected by the assigned worker is much higher (which brings the system performance down). The PSC-platform tries to specify a higher price to tasks in "difficult" regions to tempt some workers, already in the region or in its proximity, to perform the task and thereby increase the system performance.

Note that the "region" specified in [1] and for the SC-platform differs from the regions specified here and for the PSC-platform. The SC-region is defined to be the proximity (disc of certain radius, or rectangular area of certain dimensions) around each worker, where the worker will consider performing an assigned task whose location is inside it (otherwise, it will reject the task). However, the PSC-platform introduces the novel attribute of the regions for each task, in order to improve the performance of the system by 1) giving the option of performing a task inside a wider region instead of only exact location(s) and 2) assigning proportional reward for tasks that need to be performed in regions where the possibility of not being able to find appropriate workers is large.

### 2.3.8 PT-priority score

$0 \leq TaskPriorityScore(PT, PTO) \leq 1$ is the actual priority associated with each task, PT, which can help the PSC-system decide the order of the tasks which need to be processed. The higher the priority of each task is, the sooner it will be processed by the server for a proper worker assigned to it. Note that it depends on both the task, PT, and its task owner PTO. A possible formula for computing it is given by:

$$TaskPriorityScore(PT, PTO) \\ = PTOPriorityScore(PTO) \\ * TaskCatPriority(TaskCat(PT)) \\ * PTEnteredPriorityScore(PT) \\ * PTEnteredReward(PT) \\ / TaskCatReward(TaskCat(PT))$$

The task priority score is positively correlated with the priority of the PTO itself, (i.e., the higher the priority score for each task owner $PTOPriorityScore(PTO)$, the higher the priority of the tasks he/she assigns). Conversely, PTOs can decide to give higher priority scores to their requested tasks, in order to make sure that the system will treat their tasks with higher priority and assign suitable workers to them as soon as possible. Of course, in return for its special treatment of a high-priority task request by a PTO, the expectation of the system is higher earnings (received by the reward given by the PTO requesting a high priority task). Therefore, the PSC-system would check the reward suggested by the PTO to see if it is much higher than the typical reward for the given task. If so, it might consider it. Otherwise, and if the reward is almost the same, it might not consider it for a high priority treatment. (High priority requests require an on-line task-assignment, and the more such tasks, the more the demand on the CPU-time and consequently the less efficient the task assignment, and therefore, the lower the performance of the entire PSC-system.)

Moreover, the PSC-system would give a higher priority to some certain tasks depending on their task category, $TaskCat(PT)$. For example, driving a person to the airport on time might have higher priority than delivering a flower to a specific location. The reason is that any minor delay in the task of driving a person to the airport can cause the person to miss his flight, making the task a failure. However, a minor delay in the flower delivery will usually not have severe consequences and the task could still be considered a success. A PSC-system with higher number of the completed tasks before their expiration time would consider having a higher performance.

### 2.3.9 PTO Reward

$PTOReward(PT)$ is a positive real number (in dollars or points) indicating the minimum amount that the PTO needs to pay to a worker to perform and complete a requested task. The PSC-system may not accept to process a requested task from the PTO, if the $PTOReward(PT)$ is less than the typical reward value of tasks with the same category, $TaskCat(PT)$, and if it is not consistent with the strategic value of the region $Region(PT)$ where the task needs to be done. $PTOReward(PT)$ is entered by the PTO at the time the request PT is put into the system.

## 2.4 PSC-Worker

A worker in the PSC-system is a registered user of the on-line PSC service provider system who is interested in performing some certain tasks, and in some certain regions, in return for some certain reward. At the time of registration, the PSC-workers (henceforth PWs), in addition to their personal information and the minimum security and credibility questions, get to specify the task categories or the type of work that they are interested in performing as well as the correspondence expected reward for each task. Moreover, the PWs could specify the regions that they visit in their daily routines (i.e., the spatial and temporal patterns), and perhaps the other further regions that they might visit occasionally and their level of comfort to perform tasks in various regions. In addition, they might define some kind of extra compensations (beyond their typical compensation rate), in order to perform tasks in regions, times, or even types of tasks that they are not comfortable to accept at their regular rates. To make the above discussion more explicit, we mention the attributes that go with the PWs.

### 2.4.1 PW-Personal Information and id

The personal information of any registered PSC-worker along with an id associate to each worker will be archived in the PSC-system. $id(PW)$ is an integer, distinct for each user, and assigned by the system as the PW's id-number.

### 2.4.2 PW-Spatio-Temporal-Pattern

$SpatioTemporalPattern(PW, t) \subset R^2$ is the expected region where the worker PW will be at time $t$. The PSC-workers are encouraged to specify this information during registration. They may either enter exact location information or, if they are concerned with their privacy, they may specify a broader region. Note that contrary to [5], the workers do not require to disclose their exact locations even to the servers and therefore, they could keep their privacy much more secured. Being sensitive to users' privacy should encourage more workers to join the PSC system, thereby increasing its performance.

$SpatioTemporalPattern(PW, t)$ is used to represent those regions or locations that a PSC-worker (PW) is expected to visit in its daily routine. An example of the spatio-temporal pattern of an employed PW could be: 1) From Monday to Friday, 7AM to 8AM, he will be driving to work through some specific-route. 2) From 8AM to noon, he will be at work in a specified location. 3) From noon to 1PM-5PM he will be at work. 5) From 5PM-6PM he will be in some specific route driving from work to home. 6) From 6PM to 7AM he will be at some specific location at home. Of course, during weekends the expected spatio-temporal pattern will be different.

The spatio-temporal pattern of each PW will be critical in appropriately assigning workers to task. It allows the system to determine the expected distance between each PW and each task PT at any desired time $t$. If $SpatioTemporalPattern(PW, t)$ is an exact location, then the expected distance is simply the Euclidean distance between the location PW and PT; if it is a region, then the expected distance is computed from the centroid of the region.

Considering the spatio-temporal pattern of the workers is one of the key distinguishing features of the PSC system compared to earlier SC systems. It allows the system to predict the location of the workers in future times and therefore can perform the assignment of tasks to users ahead of time and during off-peak hours where the system load is light and the CPU can focus on such computations. During peak hours the CPU can therefore focus on assigning the much smaller number of tasks that arrive in real-time, and on reassigning those tasks that the assigned workers declined to perform.

Earlier SC systems that do not take spatio-temporal patterns into account must do all the assignment computations in real-time (as a result of which the CPU becomes a bottleneck and the assignment strategy suboptimal) and with a limited and local view of the system. It is therefore fully expected that considering spatio-temporal patterns will significantly boost system performance.

### 2.4.3 PW-Status

$Status(PW, t) \in [0,1]$ represents the availability of worker PW during time $t$. Workers are encouraged to enter this information when they register in the system. Here 1 represents fully available, and 0 represents unavailable. Values in between can represent the probability that PW is available at the corresponding times.

In the example, given in section 2.4.2 the PW-Status during weekdays could be: 1) $Status(PW, 7AM - 8AM) = 0.9$, 2) $Status(PW, 8AM - noon) = 0.1$, 3) $Status(PW, noon - 1PM) = 0.6$, 4) $Status(PW, 1PM - 5PM) = 0.1$, 5) $Status(PW, 5PM - 6PM) = 0.9$, and 6) $Status(PW, 6PM - 7AM) = 0.05$. On weekends it could simply be 0.

It should be clear that the availability status of a worker at any given time should be critical in determining appropriate task assignments.

### 2.4.4 PW-Task Reward

$PWReward(PW, TaskCat(PT), Region(PT))$, is the minimum reward (in dollars or points) that worker PW will accept, to perform a task in category $TaskCat(PT)$ and region $Region(PT)$. The PSC-workers are encouraged to specify this information during registration. $PWReward(PW, TaskCat(PT), Region(PT))$ can be used to specify a worker's interest in different task categories (for example, the reward a PW may ask for a delivery might be higher than the reward he asks for taking a picture). At the same time, the system may also recommend rewards for different task categories (based on what appears reasonable to other task owners and workers). But of course whether to accept these recommendations or not is up to the PWs.

Clearly, when making task assignments it is important that the reward of tasks assigned to workers meet or exceed that workers cost for the corresponding task category.

### 2.4.5 PW-Trustworthy score

$TrustworthyScore(PW, TaskCat(PT)) \in [0,1]$, is a category, represents how trustworthy worker PW is in performing a task of category $TaskCat(PT)$. Here 1 represents fully capable of performing the task and 0 represents incapable of doing so. Values in between can be interpreted as the probability that worker PW can perform as task of category $TaskCat(PT)$. Initially, the value of $TrustworthyScore(PW, TaskCat(PT))$ can be entered by the worker PW upon registration (based on how comfortable they feel with tasks of category $TaskCat(PT)$). However, as time progresses, it is important to have a dynamic trustworthy score that depends on the workers behavior. An appropriate formula could be:

$TrustworthyScore(PW, TaskCat(PT)) =$

$(m1 \frac{\# \text{ of accepted tasks } TaskCat(PT)}{\# \text{ of assigned tasks } TaskCat(PT)}$

$+ m2 \frac{\# \text{ of completed tasks } TaskCat(PT)}{\# \text{ of accepted tasks } TaskCat(PT)})/(m1 + m2)$   where m1<m2

The ratio of the number of accepted tasks to the number of assigned tasks represents how comfortable worker PW feels with tasks of category $TaskCat(PT)$. The ratio of completed tasks to accepted tasks represents the track record of worker PW with tasks of category $TaskCat(PT)$. We have taken the mean of these to be the trustworthiness of worker PW with task $TaskCat(PT)$. However, we wanted to give more weight to those who have accepted the task and completed it successfully, to those who have tried the task, instead of rejecting it (therefore, they feel somehow comfortable to perform the task). The more the ones accept the tasks without being able to perform it successfully, the lower their Trustworthy score will be, which is an indication of not being a good choice to be considered for such tasks in our future task assignments.

Even though in the SC-platform [4], there is the notion of assigning a trustworthy score to each worker in order to increase the performance of the system, however, the trustworthy score considered there is independent of the task category. This is certainly a shortcoming since certain workers may be more trustworthy with certain tasks than with others and there is the possibility of assigning a task to a worker who is incapable of doing it. Take, for example, PWA and PWB. PWA has been assigned several successful tasks before which were mostly taking pictures from some locations in her neighborhood. However, she does not drive a car. PWB, drives a car, but has fewer successful tasks in her history in the SC-system. If the new task is to deliver a flower bouquet from one location to another, based on the logic in [4], the PWA will be given the task. However, PWB is clearly the more appropriate assignee.

## 3. The Task-Assignment Algorithm

The task-assignment algorithm operates under two phases. One is off-line and during off-peak hours. In this phase, the system schedules and assigns the earlier submitted and archived tasks to appropriate workers in the system based upon the information provided on the tasks and the information available about the workers. This phase performs off-line and global optimization (which is feasible in off-peak hours when the system is not busy).

The second phase is on-line and during peak hours. In this phase, the system assigns tasks that arrive and need to be performed in short order, as well as reassigns tasks that have been rejected by workers for which they were originally assigned. The assignment in this phase is on-line and greedy and must be much less demanding of CPU time.

## 3.1 Off-Line Task Assignment

The off-line task assignment is based upon optimizing the score of assigning task PT to worker PW at time $t$. Let us begin with defining this score.

We first need to determine the time it will take for worker PW to perform task PT, if it were assigned the task at time $t$:

$$TimeToComplete(PT, PW, t)$$
$$= \|TaskLocationRegion(PT)$$
$$- SpatioTemporalPattern(PW, t)\|/v(t)$$
$$+ DurationTime(PT)$$

Here $v(t)$ is the expected velocity that a worker in the region may have at time $t$. This should be known to the system based on available expected traffic patterns. The Euclidean distance between the task PT and worker PW is based on calculating the centroids of each of their regions.

We can now define the time score for worker PW if it is assigned task PT at time t:

$$TimeScore(PT, PW, t)$$
$$= \frac{ExpirationTime(PT) - TimeToComplete(PT, PW, t) - t}{ExpirationTime(PT) - t}$$

In the above formula, the larger (closer to 1) the time score is, the less time pressure there is on the worker to finish the task before it expires. Note that a negative time score means that worker PW cannot finish task PT on time, if it is assigned the task at time $t$.

We next need to define the availability score for worker PW if it is assigned task PT at time $t$:

$$AvailabilityScore(PT, PW, t) = \frac{\int_t^{ExpirationTime(PT)} Status(PW, \tau)d\tau}{ExpirationTime(PT) - t}$$

This is simply the average availability of worker PW from time $t$ until the task's deadline.

Next we need the reward score:

$$RewardScore(PT, PW)$$
$$= \frac{Ramp(PTOReward(PT) - PWReward(TW, Cat(PT), Region(PT)))}{TaskReward(PT)}$$

Here the ramp function is the integral of the standard Heaviside function (in other words, it is zero for negative arguments and the argument itself for non-negative ones). The reward score is thus between 0 and 1 and it is zero if the reward for PT is less than the reward that PW demands for tasks that have the same category as PT.

We can now define the total score for assigning task PT to worker PW at time t as the product of the time score, the availability score, the reward score and the trustworthy score. In mathematical terms:

$$TotalScore(PT, PW, t)$$
$$= TimeScore(PT, PW, t)$$
$$* AvailabilityScore(PT, PW, t)$$
$$* RewardScore(PT, PW, t)$$
$$* TrustworthyScore(PW, TaskCat(PT))$$

We should mention that, during task assignment, we may want to give preferential treatment to task owners that have high priority (see section 2.2.1). This can be achieved by assuring that they are assigned to trustworthy workers. To do this, we can give an extra weight (based on how high a task owner's priority is $PTOPriority$) to the $TrustworthyScore$ in the above formula.

$$TotalScore(PT, PW, t)$$
$$= TimeScore(PT, PW, t) * AvailabilityScore(PT, PW, t)$$
$$* RewardScore(PT, PW, t)$$
$$* \left(TrustworthyScore(PW, TaskCat(PT))\right)^{1/PTOPriority}$$

The offline assignment algorithm is simply the one that assigns to each task PT the worker PW that maximizes the total score. In mathematical terms:

$$WorkerAssigned(PT) = \underset{PW}{argmax} \, \underset{t}{max} \, (TotalScore(PT, PW, t))$$

Similarly, the time t in which task PT will be assigned to worker PW is given by

$$TimeAssigned(PT) = \underset{t}{argmax} \, \underset{PW}{max} \, (TotalScore(PT, PW, t))$$

We should mention that if there is a conflict, in the sense that there is a single worker that maximizes the total score for more than one task, then the task it should be assigned to the one with the highest priority. If this does not break the tie then, among tasks with the highest priority, it should be assigned to the one that has the highest total score. If this does not break the tie, then it should be assigned randomly to one of the tied tasks. The remaining tasks should then be assigned to the worker that achieves the second highest score (and so on, if necessary).

Note that the algorithm just mentioned requires enumerating the total costs for all tasks, workers and times, and therefore can be time consuming. Nonetheless, it should be computationally feasible during off-hour periods. What is certainly not computationally feasible is to consider all possible *matchings* between tasks and workers, where by a matching we mean an association of *every* task to a *distinct* worker at a certain time.

Here we have taken a *task-centric* point of view and have assigned each task to a worker. It is also possible to take a *worker-centric* point of view by assigning each worker to a task (this is achieved by maximizing the total cost over PT and $t$ for any given PW). However, the latter is not very meaningful for us, since what matters is that tasks get assigned and completed, not that workers get busy by being assigned tasks.

## 3.2 On-Line Task Assignment

The off-line task assignment algorithm assigns tasks to users during the off-hour periods. While every attempt was made to find the most appropriate user for each task, there is still a chance that some users may reject their assignments at some point after receiving them. Such rejected tasks need to be reassigned. Furthermore, during peak hours, it is highly likely that some number of tasks, that need to be performed in a short time window, will also be submitted to the system. These tasks need to be assigned in a real-time on-line fashion. Here we propose a greedy algorithm that assigns tasks one at a time. Tasks are assigned in a first-come first-serve fashion, unless tasks have a higher priority (see section 2.3.8), in which case they are bumped up to the head of the queue.

### 3.2.1 On-Line Algorithm

**Input:** Task PT (for assignment or reassignment); current time $t$

**Given:** List of available workers (currently not performing other tasks)

1. For the given task PT, find the available worker that maximizes the total score at time $t$:

$$WorkerAssigned(PT) = \underset{PW}{argmax} \, TotalScore(PT, PW, t)$$

2. If the maximized total score is positive, then assign PT to the worker that maximizes it. If more than one worker does so, randomly choose one. Remove assigned worker from list of available workers and move to next PT.

3. If the maximized score is negative, it means that the task's deadline cannot be met by any of the available workers. Acknowledge failure to PTO and move to next PT.

4. If the maximized score is zero, it means that the reward is not sufficient. Query PTO to see if they are willing to increase the reward. If so, increase the award by amount determined by PTO and go to 1.

5. If not, acknowledge failure to PTO and move to next PT.

…………………………………………………………………….